\documentclass{ws-mpla}

\let\trimmarks

\def\nn{\nonumber}
\def\cite#1{[\refcite{#1}]}

\begin{document}

\markboth{Claudio A. Scrucca}
{Mediation of supersymmetry breaking in extra dimensions}

%
%

\title{\vspace{-60pt}{\footnotesize \hfill \mbox{\rm NEIP/04-10}} \\ \vspace{50pt}
MEDIATION OF SUPERSYMMETRY BREAKING\\ 
IN EXTRA DIMENSIONS}

\author{CLAUDIO A. SCRUCCA}

\address{Institut de Physique, Universit{\'e} de Neuch{\^a}tel,\\
Rue Breguet 1, CH-2000 Neuch{\^a}tel, Switzerland.\\
claudio.scrucca@unine.ch}

\maketitle


\begin{abstract}
We review the mechanisms of supersymmetry breaking mediation that occur in sequestered 
models, where the visible and the hidden sectors are separated by an extra dimension 
and communicate only via gravitational interactions. By locality, soft breaking 
terms are forbidden at the classical level and reliably computable within an effective 
field theory approach at the quantum level. We present a self-contained discussion 
of these radiative gravitational effects and the resulting pattern of soft masses,
and give an overview of realistic model building based on this set-up.
We consider both flat and warped extra dimensions, as well as the possibility that
there be localized kinetic terms for the gravitational fields.

\end{abstract}


\section{Introduction}	

The most important ingredient of supersymmetric extensions of the standard model 
is certainly the way in which supersymmetry is broken. The standard set-up consists 
of a visible sector containing the standard matter fields and a hidden sector hosting 
spontaneous supersymmetry breaking. The effect of supersymmetry breaking in the 
visible sector is then parametrized by a finite set of soft breaking terms. These soft 
terms fully characterize the phenomenology of such a model, and are for this reason 
often treated as free empirical parameters. However, it is an important and challenging 
question to find a satisfactory and natural microscopic realization of the interesting 
regions in the space of these parameters.

A particularly interesting and natural possibility is that the visible and the hidden 
sectors interact only through gravitational interactions and supersymmetry breaking 
is gravity-mediated \cite{gravmed,gravmedunif}. In this situation, the typical scale of 
the soft masses is given by $m_{\rm soft} \sim M_{\rm susy}^2/M_{\rm P}$, where 
$M_{\rm susy}$ is the scale at which spontaneous supersymmetry breaking occurs 
in the hidden sector and $M_{\rm P}$ is the Planck mass characterizing the strength 
of gravitational interactions. Since gravitational interactions are non-renormalizable, 
the soft terms depend on the unspecified UV completion of the theory. The best we 
can do in general is then to parametrize our ignorance about UV effects through 
higher-dimensional operators suppressed by suitable powers of $M_{\rm P}$, within 
a low-energy effective field theory approach. The dimensionless coefficients controlling 
these operators are expected to be generically of order one, although their precise 
values remain out of reach.

The pattern of soft terms predicted by the above scenario is qualitatively compatible with 
phenomenological requirements. In particular, all the soft masses have the same order of 
magnitude, which can be tuned to the electroweak scale by choosing $M_{\rm susy}$ to be an 
intermediate scale. However, at a more quantitative level there are some conceptual problems.
Most importantly, the scalar soft masses must be nearly flavour-universal in order not to spoil 
the Glashow--Iliopoulos--Maiani mechanism suppressing flavour-changing processes in the 
standard model. There is however no obvious reason for this to be the case in general, even 
assuming a theoretical description of the flavour structure of the theory relying on some 
new symmetry that is spontaneously broken at a scale $M_{\rm flav}$. Indeed, since 
$M_{\rm P}$ is supposed to be close to the fundamental scale of the theory, it is natural 
to imagine that $M_{\rm flav}$ is of the same order of magnitude as it, 
and the soft terms are thus expected to be flavour-generic.

The essence of the above-described supersymmetric flavour problem lies in the fact 
that even if the soft scalar masses are taken to be flavour-diagonal at the classical level, 
quantum effects will generically induce unacceptably large flavour-breaking corrections. 
A natural solution to this problem could then come from some mechanism that is able to screen the 
soft scalar masses from these dangerous effects. However, it is straightforward to convince one-self 
that it is impossible to implement this by just imposing some kind of approximate symmetry.
Interestingly enough, a radically new possibility turns out to emerge from the notion of locality
in the presence of extra dimensions \cite{anomed1}. The crucial idea is that if the visible and 
the hidden sectors are separated by an extra dimension, any transmission effect for supersymmetry 
breaking must necessarily correspond to operators that are non-local along the internal dimensions.
As a consequence, the soft scalar masses will vanish at the classical level and be dominated 
by soft quantum effects that are flavour-universal and saturated in the IR, at the compactification 
scale. These effects are finite and depend in a significant way only on the spectrum 
of light modes of the higher-dimensional theory. They can thus be reliably computed within an 
effective supergravity description.

\section{Supersymmetry breaking and sequestering}

Let us start by describing the way in which supersymmetry is broken in gravity-mediated 
models. To do so, we consider a supergravity theory with visible and hidden sectors 
labelled by the index $i={\rm v},{\rm h}$. The minimal set-up is obtained by introducing some sample 
chiral and vector multiplets $\Phi_i=(\phi_i,\chi_i; F_i)$ and $V_i=(A^\mu_i,\lambda_i;D_i)$
in these two matter sectors. Using the superconformal approach to supergravity 
\cite{supconf}, the gravitational interactions are described by a conformal gravity multiplet 
$C = (e_\mu^a,\psi_\mu;a_\mu,b_\mu)$ and a conformal compensator chiral multiplet 
$S=(\phi_S,\psi_S;F_S)$. The full superconformal group is gauged and is reduced to the 
ordinary super-Poincarr{\'e} group by gauge-fixing the extra local symmetries through the 
conditions $b_\mu = 0$, $\phi_S = 1$ and $\psi_S = 0$.

Using the above framework, the couplings between the matter fields and the scalar 
auxiliary field of supergravity that are relevant for supersymmetry breaking can be written 
in a particularly simple and illuminating form. Indeed, the general form of this part of the Lagrangian 
is obtained by taking ordinary rigid supersymmetry $D$ and $F$ densities of an arbitrary K\"ahler 
function $\Omega$ and an arbitrary superpotential and kinetic functions $W$ and $G_i$.
At the classical level of the effective theory, the dependence on the conformal compensator 
is uniquely fixed by Weyl invariance and the matter Lagrangian reads:
\begin{eqnarray}
{\cal L}_{\rm mat} &=& 
\big[\Omega(\Phi_j,\Phi_j^\dagger) S S^\dagger\big]_D 
+ \big[W(\Phi_j) S^3 \big]_F 
+ \big[W(\Phi_j) S^3 \big]_F^\dagger \nn \\
&\;& + \big[G_i(\Phi_j) {\cal W}_i^2]_F 
+ \big[G_i(\Phi_j) {\cal W}_i^2 \big]_F^\dagger \;.
\end{eqnarray}
The functions $\Omega$, $W$ and $G_i$, which parametrize this Lagrangian, have 
general expressions consisting of infinite series in inverse powers of $M_{\rm P}$:
\begin{eqnarray}
\Omega &=& - 3 M_{\rm P}^2 + \Phi_{\rm v} \Phi_{\rm v}^\dagger + \Phi_{\rm h} \Phi_{\rm h}^\dagger
+ \frac h{M_{\rm P}^2}\, \Phi_{\rm v} \Phi_{\rm v}^\dagger \Phi_{\rm h} \Phi_{\rm h}^\dagger
+ \cdots \;, \label{Omegagen} \\
W &=& \Lambda^3 + M_{\rm susy}^2 \Phi_{\rm h} 
+ \cdots \;, \label{Pgen}\\
G_i &=& g_{i}^{-2} + \delta_{i,h} \frac k{M_{\rm P}}\, \Phi_{\rm h}
+ \cdots \label{Ggen} \;. 
\end{eqnarray}
For simplicity, we have discarded all the dependence on the gauge fields other than 
their quadratic kinetic terms, and left the gauge quantum numbers of the matter fields
unspecified. In the above schematic writing, it should be remembered that the second,
third and fourth terms in (\ref{Omegagen}) are always admissible, whereas the second 
terms in (\ref{Pgen}) and (\ref{Ggen}) can occur only for singlets. Moreover, in order to 
achieve a vanishing cosmological constant, we must in general tune 
$\Lambda^3 \sim M_{\rm susy}^2 M_{\rm P}$. The supersymmetry breaking VEVs of the 
auxiliary fields are then found to be:
$$
F_{\rm h} \sim M_{\rm susy}^2 \;,\;\; 
F_{S} \sim \frac {\Lambda^3}{M_{\rm P}^2} \sim \frac {M_{\rm susy}^2}{M_{\rm P}} \;.
$$
It follows that the soft masses are given by
\begin{eqnarray}
&& m_{3/2} \sim |F_{S}| \sim \frac {M_{\rm susy}^2}{M_{\rm P}} \;, \\
&& m_{1/2} \sim k\, \frac {|F_{\rm h}|}{M_{\rm P}}  
\sim k\, \frac {M_{\rm susy}^2}{M_{\rm P}} \;, \\
&& m_{0}^2 \sim h\, \frac {|F_{\rm h}|^2}{M_{\rm P}^2}  
\sim h\, \frac {M_{\rm susy}^4}{M_{\rm P}^2} \;.
\end{eqnarray}
From the above expressions, we see that the soft gaugino and scalar masses are controlled
by the dimensionless coefficients $h$ and $k$ of the leading operators mixing the two sectors.
Their values are undetermined within the effective theory, and are therefore a priori completely generic.
In particular, the parameter $h$ can have an arbitrary sign and flavour structure.

At the quantum level, the low-energy effective theory can acquire an anomalous dependence on 
the compensator multiplet due to the conformal anomaly induced by quantum fluctuations 
of the light gauge fields. More precisely, Weyl invariance dictates that the dependence on the 
energy scale $\mu$ of all the running couplings must come together with a dependence on the 
compensator multiplet $S$, in such a way that only the quantity $\mu/\sqrt{S S^\dagger}$ appears.
This leads to anomaly-mediated corrections to the soft masses. In terms of the gauge loop factor 
$\alpha_{\rm gau} = g_0^2/(16 \pi^2)$, they have the form \cite{anomed1,anomed2}
\begin{eqnarray}
&& \delta m_{1/2} \sim a\,\alpha_{\rm gau}\, |F_S| 
\sim a\,\alpha_{\rm gau}\, \frac {M_{\rm susy}^2}{M_{\rm P}} \;, \\
&& \delta m_{0}^2 \sim b\,\alpha_{\rm gau}^2\, |F_S|^2  
\sim b\,\alpha_{\rm gau}^2\, \frac {M_{\rm susy}^4}{M_{\rm P}^2} \;.
\end{eqnarray}
The coefficients $a$ and $b$ are group-theoretical factors that depend only the quantum 
numbers of the corresponding fields. In particular, the coefficient $b$ is flavour-universal 
and has a definite sign, which turns out to be positive for asymptotically free gauge groups 
and negative otherwise. This implies that the anomaly-mediated contribution is positive 
for the squarks and negative for the sleptons.

In a generic situation, the coefficients controlling the gravity-mediated contributions to the 
soft masses are of order one, whereas the coefficients controlling the anomaly-mediated 
contributions are instead of order $\alpha_{\rm gau}$. The latter represent thus negligible 
corrections and the situation is as bad as described in the introduction.
This matter of state does however radically change in sequestered models. Indeed, as 
already explained, in such models the dimensionless coefficients $h$ and $k$ are suppressed 
and computable. They are forced to vanish at the tree-level in the counting of heavy-mode 
loops, and their values are therefore determined by the one-loop effect. These are easily estimated by simple 
power counting to be of the order of $\alpha_{\rm gra} = M_{\rm KK}^2/(16 \pi^2 M_{\rm P}^2)$, that is 
a loop factor for gravity saturated at the compactification scale $M_{\rm KK}$, which is defined as 
the mass scale of the Kaluza--Klein modes divided by $\pi$ and represents the scale at 
which the constraints put by locality become effectively significant. In sequestered models, 
we therefore get:
\begin{eqnarray}
k = c\, \alpha_{\rm gra} \;,\;\;
h = d\, \alpha_{\rm gra} \;.
\end{eqnarray}
The coefficients $c$ and $d$ are numbers of order one that can be deduced only through
an explicit computation. In particular, $d$ is flavour-universal but can a priori have an arbitrary 
sign.

From the above discussion, we understand that sequestered models can potentially solve the 
supersymmetric flavour problem of gravity-mediated models and provide a framework where 
the soft scalar masses are naturally universal and radiatively induced by two distinct effects
associated to gauge and gravitational loops. In order to get a interesting model, these two 
effects must be able to compete, in order for the slepton masses squared to have a chance 
to be positive \cite{Chacko:1999am}. This can happen if $\alpha_{\rm gra} \sim \alpha_{\rm gau}^2$, 
that is:
\begin{equation}
\frac {M_{\rm KK}}{M_{\rm P}} \sim 4 \pi \alpha_{\rm gau} \;.
\label{est}
\end{equation}
This is a reasonable possibility, which is compatible with the use of an effective low-energy 
approach and can therefore be efficiently investigated in concrete models with extra dimensions.
It gives a strong motivation for explicitly computing the above discussed gravitational loop 
contributions to the scalar soft masses in concrete classes of sequestered models and investigate 
their viability more precisely.
This program involves also studying and taking into account the additional effects occurring in 
these models due to the radion multiplet parametrizing the dynamics of the extra compact 
dimension. We will see in the following that these are qualitatively similar to those discussed 
in this section for four-dimensional models, but they will turn out to be absolutely crucial for
the possibility of having a positive contribution from the gravitational loop effects.

\section{Minimal sequestered models}

The simplest realization of sequestered models is achieved by introducing a single 
extra dimension and locating the visible and the hidden sectors at two branes positioned
at fixed points in the internal dimension. We shall consider the general case of a warped 
geometry of the Randall--Sundrum type \cite{Randall:1999ee}. 
The formulation of the locally supersymmetric version of this class of 
compactifications has been the object of active study in the past years. 
\footnote{There are two different formulations, developed respectively in 
refs.~\cite{RSwarped1,BKV} and \cite{RSwarped2,Zucker}, which differ by a singular 
local $R$-symmetry transformation \cite{RS12equiv}. 
We use here the first.} It consists in a gauged 5D supergravity theory compactified 
on the orbifold $S^1/{\bf Z}_2$. We denote by $y$ the coordinate of the internal circle 
and by $R$ its radius. The two fixed points of the orbifold action are located at 
$y_0 = 0$ and $y_1 = \pi R$, and they represent the boundaries of the physical 
segment of internal space. Using the notation $\delta_i(y) = \delta(y-y_i)$,
the Lagrangian of the theory takes then the form:
\begin{eqnarray}
{\cal L} = {\cal L}_5 + \delta_0(y) {\cal L}_0 + \delta_1(y) {\cal L}_1 \;.
\label{Lagrangian}
\end{eqnarray}
The kinetic parts of the bulk and boundary Lagrangians are given by 
\begin{eqnarray}
{\cal L}_5^{\rm kin} &=& \sqrt{g_5} \Big[\!\!-\! \Lambda_5\! 
- \frac 12 M_5^3 \Big({\cal R}_5 \!+ i \bar \Psi_{\!M} \Big(\Gamma^{MRN}\! D_R 
\!-\! \frac {3i}2 k \epsilon(y) \Gamma^{MN}\Big)\Psi_{\!N} 
\!+ \frac 12 F_{MN}^2 \Big) \Big]\,,\;\quad \\
{\cal L}_i^{\rm kin} &=& \sqrt{g_4} \Big[\!\!-\! \Lambda_i - \frac 12 M_i^2 \Big({\cal R}_4 
+ i \bar \Psi_\mu \gamma^{\mu\rho\nu} D_\rho \Psi_\nu \Big) 
+ \Big(\!-|\partial_\mu \phi_i|^2 + i \bar \psi_i \gamma^\mu D_\mu \psi_i \Big) 
\Big] \,.\; \quad
\end{eqnarray}
In these expressions $M_5$ is the 5D fundamental energy scale, $k$ is a curvature scale 
and $M_{0,1}$ are two scales parametrizing possible localized kinetic terms for the bulk 
fields, whereas $\Lambda_5$ and $\Lambda_{0,1}$ are bulk and boundary cosmological 
constants that are tuned to the values
$\Lambda_5 = - 6 M_5^3 k^2$ and $\Lambda_0 = - \Lambda_1 = 6 M_5^3 k$.

The above theory has a non-trivial supersymmetric warped solution defining a 
slice of an ${\rm AdS}_5$ space that is delimited by the two 4D branes at $y=y_{0,1}$. 
Defining the function $\sigma(y) = k |y|$, the background is given by  \cite{Randall:1999ee}:
\begin{equation}
g_{MN} = e^{-2 \sigma} \eta_{\mu \nu} \delta^\mu_M \delta^\nu_N 
+ \delta^y_M \delta^y_N \;,\;\; \Psi_M = 0 \;,\;\; A_M = 0 \;.
\label{background}
\end{equation}
At energies much below the compactification scale $M_{\rm KK}$, the fluctuations 
around the above background solution are described by an ordinary ungauged 4D 
supergravity theory with vanishing cosmological constant. The fields of this effective 
theory are the massless zero modes of the 5D theory and fill out a supergravity multiplet 
$G = (h_{\mu \nu},\psi_\mu)$ and a radion chiral multiplet $T = (t, \psi_t)$ of 
the on-shell surviving supersymmetry. They are defined by parametrizing the 
fluctuations around the background as follows:
\begin{eqnarray}
g_{\mu\nu} &=& \exp \!\Big(\!\!-\!2 \sigma \frac {{\rm Re}\,t}{\pi R}\Big) 
\big(\eta_{\mu \nu} + h_{\mu \nu} \big) \;,\;\;
g_{\mu y} = 0 \;,\;\; g_{yy} = \Big(\frac {{\rm Re}\,t}{\pi R}\Big)^2 \;; \nn \\
\Psi_\mu &=& \psi_\mu \;,\;\; \Psi_y = \frac {\sqrt{2}}{\pi}\psi_t \;;\;\;
A_\mu = 0 \;,\;\; A_y = \frac {\sqrt{3}}{\sqrt{2}\pi}\,{\rm Im}\,t\;.
\label{fluctuations}
\end{eqnarray}
The relevant low-energy dynamics is then fully controlled by the K\"ahler
function and the superpotential of the effective theory as a function of 
the radion multiplet $T$ and the matter multiplets $\Phi_i$. These are
computed by integrating out the heavy Kaluza--Klein modes of the bulk fields.

At the classical level, the effective K\"ahler function can be deduced 
by substituting eqs.~(\ref{fluctuations}) into eq.~(\ref{Lagrangian}) and 
integrating over the internal dimension. The result has the form 
\cite{effactionRS,stabwarp1}:
\begin{eqnarray}
\Omega(T\!+\!T^\dagger,\Phi_i,\Phi_i^\dagger) &=&
- 3 \frac {M_5^3}{k} \Big(1 - e^{- k (T + T^\dagger)} \Big) 
\nn \\ &\;& 
+\, \Omega_0(\Phi_0,\Phi_0^\dagger) 
+ \Omega_1(\Phi_1,\Phi_1^\dagger) \,e^{- k (T + T^\dagger)} \;.\qquad
\label{kahlereff}
\end{eqnarray}
The first term is the matter-independent contribution from the bulk, whereas 
the last two terms are the matter-dependent contributions from the branes.
Ignoring irrelevant operators involving higher powers of $\Phi_i \Phi_i^\dagger$,
the latter have the form
\begin{equation}
\Omega_i^{\rm kin}(\Phi_i,\Phi_i^\dagger) = - 3 M_i^2 + \Phi_i \Phi_i^\dagger \;.
\label{kahlerboundary}
\end{equation}
Notice that whereas $M_0$ is arbitrary, $M_1$ must be smaller than $\sqrt{M_5^3/k}$;
otherwise, the radion scalar field parametrized by $e^{-k T}$ would become a ghost \cite{Luty:2003vm}. 
Moreover, when these parameters come close to their maximal values, the effective 
fundamental scale of the theory is significantly lowered \cite{Luty:2003vm}. 
The effective superpotential, on the other hand, is specified by 
the superpotentials $W_i$ localized on the branes:
\begin{eqnarray}
W(T,\Phi_i) = W_0(\Phi_0) + W_1(\Phi_1)\, e^{- k T} \;.
\label{supereff}
\end{eqnarray}
The effective Planck scale can be read off from (\ref{kahlereff}). Assuming that the 
matter fields have vanishing VEVs and recalling that $T$ has VEV $\pi R$, one finds:
\begin{equation}
M_{\rm P}^2 = \frac {M_5^3}{k} \Big(1 - e^{- 2 \pi k R} \Big) 
+ M_0^2 + M_1^2\, e^{- 2 \pi k R} \;.
\end{equation}
The scalar soft masses vanish, independently of which fixed points are chosen to host the visible and 
the hidden sectors. If ${\rm v}=0$ and ${\rm h}=1$, this is obvious, since the visible sector multiplet $\Phi_0$ 
does not couple neither to $T$ nor to $\Phi_1$. If ${\rm v}=1$ and ${\rm h}=0$, the visible sector multiplet 
$\Phi_1$ couples instead to $T$, although not to $\Phi_0$, in the K\"ahler function. But it does 
in a very particular way, as a conformal compensator, which can be trivialized through a field 
redefinition, and this guarantees that all the masses cancel. The classical effective theory is 
thus of the sequestered type.

At the quantum level, there occur two kinds of corrections to the K\"ahler function.
The first class represents a trivial renormalization of the local operators corresponding 
to the classical expression (\ref{kahlereff}). These UV effects are divergent and 
incalculable, but anyhow irrelevant, since as explained above they cannot induce 
soft masses. A second class corresponds instead to new effects that have a field 
dependence that differs from the one implied by locality and general covariance in 
(\ref{kahlereff}). These IR effects are finite and calculable, and control relevant 
contributions to the soft masses. The relevant corrections that we have to compute 
are therefore non-local and radion-dependent, and can be parametrized in the following 
general form:
\begin{equation}
\Delta \Omega(T\!+\!T^\dagger,\Phi_i,\Phi_i^\dagger) = 
\sum_{n_0,n_1=0}^\infty C_{n_0,n_1}(T \!+\! T^\dagger) (\Phi_0 \Phi_0^\dagger)^{n_0} 
(\Phi_1 \Phi_1^\dagger)^{n_1} \;.
\label{corrkahlereff}
\end{equation}
The functions $C_{n_0,n_1}$ control the leading effects allowing the transmission 
of supersymmetry breaking from one sector to the other. More precisely, supersymmetry
breaking induces in general non-vanishing values for the auxiliary fields of all the non-visible
chiral multiplets in the theory, that is the compensator, the radion multiplet and the hidden sector 
matter multiplets. The operator associated with $C_{0,0}$ yields then a Casimir energy,
the one associated to $C_{1,0}$ or $C_{0,1}$, depending one where the visible and the hidden 
sectors are put, a radion-mediated scalar mass squared, and finally the one associated 
to $C_{1,1}$ a brane-mediated scalar mass squared. In the flat limit of vanishing warping, 
$C_{0,0}$ was computed in ref.~\cite{Ponton:2001hq}, $C_{1,0}$ and $C_{0,1}$ first in 
\cite{Gherghetta:2001sa} for $M_i = 0$ and then in \cite{Rattazzi:2003rj} for $M_i \neq 0$, 
and finally $C_{1,1}$ in \cite{Rattazzi:2003rj,Buchbinder:2003qu} for $M_i = 0$ and 
in \cite{Rattazzi:2003rj} for $M_i \neq 0$. In the general case of finite warping, the first coefficient 
was computed in refs.~\cite{Casimirwarped} (see also \cite{Casimirwarped2}) and the other 
ones in \cite{Gregoire:2004nn}.

\section{One-loop corrections}  

There are various techniques that can be used to perform the computation of the above described 
gravitational quantum corrections. A first approach, which was taken in ref.~\cite{Rattazzi:2003rj},
consists in using the off-shell component description developed in ref.~\cite{Zucker} with the 
same philosophy as ref.~\cite{Mirabelli:1998aj}, and focus on a particular component of 
each operator. A second approach, which was taken in refs.~\cite{Buchbinder:2003qu} 
and \cite{Gregoire:2004nn}, consists instead in using the linearized superfield approach of
ref.~\cite{Linch:2002wg} or the generalization to the warped case of \cite{Gregoire:2004nn},
and perform a supergraph computation. But happily, after some supergravity gymnastics the 
corrections we are interested can be mapped to the effective action of a very simple theory involving 
a single real scalar degree of freedom propagating in the same geometry and possessing localized kinetic 
terms at the two branes. We will not give here a rigourous proof of this fact, for which we refer the interested 
reader to ref.~\cite{Gregoire:2004nn}, but rather some qualitative hints on the basic properties behind it.

The main ingredients of the computation are the interactions between the brane 
and the bulk fields. These are fixed by the local symmetries that the theory 
should possess at the branes, and are most conveniently organized by splitting
the bulk fields in background values plus fluctuations, and expanding in powers
of the latter. The leading term is given by the minimal coupling between the 
supercurrent of the matter theory and the supergravity fluctuation multiplet,
and leads to cubic vertices involving two brane and one bulk fields. At the next 
order, we find instead a coupling between the K\"ahler function of the matter 
theory and the kinetic term of the supergravity fluctuation, which leads to 
quartic vertices with two brane and two bulk fields. The one-loop diagrams that 
contribute to each operator in eq.~(\ref{corrkahlereff}), with 
given $n_0$ and $n_1$, are obtained by combining these two types of vertices 
and fall into two classes. The first class consists of all those diagrams that 
involve some cubic vertices. Fortunately, although individually gauge-dependent, 
these turn out to sum up to zero in any gauge. The second class consists of the 
unique diagram that is made of only quartic vertices. This is manifestly 
gauge-independent and represents the only net contribution. Moreover, it is particularly 
simple to compute, since the involved interactions have the form of additional kinetic 
terms for the bulk fields with matter-dependent coefficients and localized at the two branes.

Proceeding along the above lines, and using a linearized superfield formulation 
of the theory, it is possible to compute the full correction (\ref{corrkahlereff}) 
in closed form. The first step is to realize that the kinematical structure
of the correction, namely those factors that depend on the tensor and superspace 
structure of the virtual particles and their interactions, amount to a universal 
factor of $-4/p^2$, where $p$ is the internal momentum in the loop. The factor $4$ 
takes into account the multiplicity of bosonic and fermionic degrees of freedom and 
the factor $1/p^2$ the fact that the K\"ahler function determines the component 
effective action only after taking its $D$ component. The dynamical part of the 
correction that is left over can then be described in terms of a single real scalar 
degree of freedom $\varphi$ propagating in the same background geometry and having 
some localized kinetic terms $l_i$, with Lagrangian
\begin{equation}
{\cal L} = -\frac 12  e^{-2\sigma(y)} \Big[(\partial_\mu \varphi)^2 
+ e^{-2\sigma(y)}(\partial_y \varphi)^2
+ \Big(l_0 \delta_0(y) + l_1 \delta_1(y) \Big) 
(\partial_\mu \varphi)^2 \Big] \;.
\label{Lagphi}
\end{equation}
More precisely, the one-loop correction to the K\"ahler function is obtained 
from the one-loop effective action for the above theory by inserting a factor $-4/p^2$ 
in its virtual momentum representation, and promoting the parameters $R$ and $l_i$ 
on which it depends to the superfields $(T + T^\dagger)/(2\pi)$ and 
$-\Omega_i^{\rm kin}/(3 M_5^3)$ respectively. 
\footnote{The fact that a displacements of each brane is effectively equivalent to a flow in the local 
operators on it \cite{Lewandowski:2002rf} can be used to relate to some extent the matter 
dependence to the radion dependence. It allows to derive the former from the latter at 
leading order and under some restrictions \cite{Gregoire:2004nn}.}

A convenient way to compute the effective action for the scalar theory defined by the 
Lagrangian (\ref{Lagphi}) is to start with $l_i=0$ and to reconstruct the full result 
for $l_i \neq 0$ by resumming all the diagrams with $l_i$ insertions.  
We thus denote by $\Psi_n(y)$ and $m_n$ the wave functions and the masses of the 
Kaluza--Klein modes of the scalar field $\varphi$ in the limit $l_i=0$. The building blocks for the 
computation of the matter-dependent terms are the boundary-to-boundary propagators 
connecting the points $y_i$ and $y_j$, with $y_{i,j}=0,\pi R$. More precisely, for 
later convenience we include a suitable power of the induced metric at the two points
that the propagator connects, and consider the quantities
\begin{equation}
\Delta_{ij}(p) = \sum_n e^{-\frac 32 \sigma(y_i)} e^{-\frac 32 \sigma(y_j)} 
\frac {\Psi_n(y_i) \Psi_n(y_j)}{p^2 + m_n^2} \;.
\label{defprop}
\end{equation}
The quantity that is relevant to compute the matter-independent term is instead the 
following spectral function:
\begin{equation}
Z(p) = \prod_n \big(p^2 + m_n^2\big) \;.
\label{E}
\end{equation}

The above quantities are difficult to evaluate in terms of their definitions as infinite 
sums and products over the Kaluza--Klein spectrum, because there exist no simple closed form 
expressions for $\Psi_n(y)$ and $m_n$. Fortunately, they can be deduced 
in a much simpler way as solutions to differential equations. The results are most 
conveniently written in terms of the functions $\hat I_{1,2}$ and $\hat K_{1,2}$, defined 
in terms of the standard Bessel functions $I_{1,2}$ and $K_{1,2}$ as
\begin{equation}
\hat I_{1,2}(x) = \sqrt{\frac {\pi}{2}} \sqrt{x}\,I_{1,2}(x) \;,\;\;
\hat K_{1,2}(x) = \sqrt{\frac {2}{\pi}} \sqrt{x}\,K_{1,2}(x) \;.
\label{hatIK}
\end{equation}

Consider first the quantities (\ref{defprop}). To compute them, we exploit the fact that 
they are connected by the simple relation 
$\Delta_{ij}(p) = e^{-\frac 32 \sigma(y_i)} e^{-\frac 32 \sigma(y_j)}\Delta(p,y_i,y_j)$
to the propagator $\Delta(p,y,y^\prime)$ of the scalar field $\varphi$ for $l_i=0$ in 
mixed momentum/position space for the non-compact/internal directions. The latter is 
defined as the solution with Neumann boundary conditions at $y$ equal to $0$ and $\pi R$ 
of the differential equation $(e^{-2ky}p^2 - \partial_y e^{-4ky} \partial_y) \Delta(p,y,y^\prime) 
= \delta(y - y^\prime)$.
This is most easily solved by switching to the conformal variable $z = e^{ky}/k$, in which the 
positions of the two branes are given by $z_0 = 1/k$ and $z_1 = e^{k \pi R}/k$. 
The result has been derived in ref.~\cite{propRS}, and when restricted to the brane positions, 
it finally yields:
\begin{eqnarray}
\Delta_{00}(p) &=& \frac {1}{2p} \frac
{\hat I_1(p z_1) \hat K_2(p z_0) + \hat K_1(p z_1) \hat I_2(p z_0)}
{\hat I_1(p z_1) \hat K_1(p z_0) - \hat K_1(p z_1) \hat I_1(p z_0)} \;, 
\label{prop00} \\
\Delta_{11}(p) &=& \frac {1}{2p} \frac
{\hat I_1(p z_0) \hat K_2(p z_1) + \hat K_1(p z_0) \hat I_2(p z_1)}
{\hat I_1(p z_1) \hat K_1(p z_0) - \hat K_1(p z_1) \hat I_1(p z_0)} \;, 
\label{prop11} \\[1mm]
\Delta_{01,10}(p) &=& \frac {1}{2p} \frac
{1}{\hat I_1(p z_1) \hat K_1(p z_0) - \hat K_1(p z_1) \hat I_1(p z_0)} \;.
\label{prop0110}
\end{eqnarray}

Consider next the formal determinant (\ref{E}). Although this is not a propagator, it can still be functionally 
related to it. Indeed, the masses $m_n$ are by definition the poles of the latter, that is the zeroes of 
$F(p) = \hat I_1(p z_1) \hat K_1(p z_0) - \hat K_1(p z_1) \hat I_1(p z_0)$.
The infinite product in eq.~(\ref{E}) has the form of an irrelevant constant divergent prefactor times a finite 
function of the momentum.  In order to compute the latter, we consider the quantity 
$\partial_p\,{\rm ln}\, Z(p) = \sum_n 2p/(p^2 + m_n^2)$. The infinite sum over the eigenvalues, defined
by $F(i m_n) = 0$, is now convergent and can be computed with standard techniques. The result is simply 
$\partial_p\,{\rm ln}\, F(p)$. This implies that $Z(p) = F(p)$, up to the already mentioned  irrelevant 
infinite overall constant:
\begin{equation}
Z(p) = \hat I_1(p z_1) \hat K_1(p z_0) - \hat K_1(p z_1) \hat I_1(p z_0) \;.
\end{equation}

Having the expressions for $\Delta_{ij}(p)$ and $Z(p)$ for $l_i=0$, 
we can compute the full effective action for $l_i \neq 0$ by summing up all 
the diagrams with an arbitrary number of insertions of these localized kinetic 
terms. The vacuum diagram without any interaction involves only the spectral 
function $Z$. All the other diagrams involve instead the propagators $\Delta_{ij}$,
and their sum reconstructs the determinant in the two-dimensional space 
of branes of the identity plus the interactions times the propagators. Since the 
interaction localized at $z_i$ involves a factor $(kz_i)^{-2}$, and we have
included a factor $(k z_i)^{-\frac 32}(kz_j)^{-\frac 32}$ in the definition of 
the $\Delta_{ij}$'s, each of the latter comes along 
with a factor $l_i k z_i$. The expression for the correction to the K\"ahler 
function is finally found by inserting a factor $-4/p^2$ in the virtual 
momentum integral, as explained above. The result reads:
\begin{eqnarray}
\label{oneloopdiv}
\Omega_{\rm 1-loop} &=& \frac 12 \! \int\! \frac{d^4p}{(2\pi)^4} \frac{-4}{p^2} \ln \bigg\{ Z(p) \det \bigg[
\mathbf{1} + kp^2 \bigg(
\begin{array}{cc}l_0 z_0 \Delta_{00}(p) \,&\, l_0 z_0 \Delta_{01}(p) \smallskip \cr
l_1 z_1 \Delta_{10}(p) \,&\, l_1 z_1 \Delta_{11}(p)
\end{array}
\bigg)\bigg]\bigg\} \;.\qquad
\label{eq:effaction}
\end{eqnarray}

The momentum integral in eq.~(\ref{eq:effaction}) is of course divergent. This corresponds to 
the fact that it also contains incalculable corrections to the coefficients
of the local operators that are already allowed to appear at the classical level.
In order to disentangle the finite corrections associated to the non-local quantities that 
we are interested in, we thus need to subtract these divergent contributions
corresponding to the renormalization of local terms. The appropriate
subtraction can be identified by replacing the brane-to-brane propagators 
$\Delta_{ij}(p)$ and the spectra function $Z(p)$ with their asymptotic behaviours 
$\tilde \Delta_{ij}(p)$ and $\tilde Z(p)$ for $p\to\infty$. Up to exponentially suppressed 
terms of order $e^{-2p(z_1-z_0)}$, which are clearly irrelevant, we find:
\begin{eqnarray}
&& \tilde\Delta_{00}(p) = \frac {1}{2p} \frac {\hat K_2(p z_0)}{\hat K_1(p z_0)} \;,\;\;
\tilde \Delta_{11}(p) = \frac {1}{2p} \frac {\hat I_2(p z_1)}{\hat I_1(p z_1)} \;,\;\;
\tilde \Delta_{01,10}(p) = 0 \;, \\
&& \tilde Z(p) = \hat I_1(p z_1) \hat K_1(p z_0) \;.
\end{eqnarray}
The divergent part of eq.~(\ref{eq:effaction}) is then obtained by replacing
each untilded quantity with its tilded limit. This yields
\begin{equation}
\label{eq:substraction}
\Omega_{\rm div}= \frac 12 \!  \int\! \frac{d^4 p}{\left(2 \pi \right)^4} \frac{-4} {p^2} 
\Big[\ln \tilde Z(p) + \raisebox{0pt}{$\sum_i$} \ln 
\left(1 + k p^2 l_i z_i \tilde \Delta_{ii}(p) \right)\Big] \;.
\end{equation}
It can be verified, by using an explicit cut-off and rescaling the integration 
variable to make the dependence of each term on the $z_i$'s explicit, that 
the above expression indeed has the structure of the most general allowed 
counterterm.

The non-local corrections to the K\"ahler function can now be computed by subtracting 
from the total one-loop expression (\ref{oneloopdiv}) the divergent contribution 
(\ref{eq:substraction}) associated to the local corrections. Our final result is 
then given by
\begin{eqnarray}
\Delta \Omega &=& \int  \frac {d^4p}{(2 \pi)^4} \frac {-2}{p^2}\, 
{\rm ln}\, \frac {Z(p)}{\tilde Z(p)} 
\frac {\prod_i\Big(1 + k z_i l_i p^2 \Delta_{ii}(p)\Big) 
- \prod_i \Big(k z_i l_i p^2 \Delta_{ii^\prime}(p)\Big)}
{\prod_i\Big(1 + k z_i l_i p^2 \tilde \Delta_{ii}(p)\Big)} \;.\qquad
\label{result}
\end{eqnarray}
As already explained, the parameters $l_i$ and $z_i$ on which this expression
depends must be promoted to superfields; defining for convenience the lengths 
$\alpha_i = M_i^2/M_5^3$, this is done according to the following rules:
\begin{equation}
l_j \to \alpha_j - \frac {\Phi_j^\dagger \Phi_j}{3 M_5^3} \;,\;\;
z_j \to \frac 1k e^{-\frac j2 k (T+T^\dagger)} \;.
\end{equation}

The functions $C_{n_0,n_1}$ can now be deduced by expanding eq.~(\ref{result}) in 
powers of the matter superfields and comparing the result with the general expression 
(\ref{corrkahlereff}). They have of course the form of $(n_0+n_1)$-point amplitudes 
involving the dressed brane-to-brane propagators and spectral function, in the presence 
of localized kinetic terms $\alpha_i$. Their functional dependence on the radion superfield 
has the structure of an infinite series in powers of $e^{- k (T+T^\dagger)}$, and can be 
studied numerically. The most important result is that the functions $C_{0,0}$, $C_{1,0}$ and 
$C_{0,1}$ are positive for $\alpha_i = 0$ but can become negative for sufficiently large 
$\alpha_i \neq 0$, the transitions points depending on the warping $k$, whereas $C_{1,1}$ 
stays always positive for any value of $\alpha_i$ and $k$. This implies that the Casimir 
energy and the radion-mediated squared masses can be either negative or positive, whereas 
the brane-mediated squared masses are always negative. An important general observation 
is that the effects of the warping and of the localized kinetic terms are locally similar,
as expected from the fact that they affect in similar ways the masses and the wave
functions of the Kaluza--Klein modes, but globally differ in crucial aspects. More precisely, 
a sign flip in the first three coefficients is possible only in the presence of a 
non-vanishing localized kinetic term. Warping only influences the size of the 
critical value of the localized kinetic that is needed for the sign flip.  This means that
to have a chance to obtain positive scalar squared masses in these minimal sequestered 
modes, one has to rely on a localized kinetic term in the hidden sector, with a size that 
depends on the warping. To be more quantitative and discuss model building, we shall 
specialize in the next sections to the two extreme cases of weak and strong warping.

\section{Weakly warped models}

In the weak warping limit $k R \ll 1$, the low-energy effective theory is most conveniently described  
by using as radion superfield $T$. The effective K\"ahler function can then be organized as
a series in powers of the small parameter $k(T\!+\!T^\dagger)$. Retaining only the leading term, the 
tree-level part is given by
\begin{equation}
\Omega = - 3 M_5^3 \big(T + T^\dagger + l_0 + l_1 \big)\;.
\end{equation}
Similarly, the one-loop correction (\ref{result}) simplifies to
\begin{equation} 
\Delta \Omega = \frac 1{4 \pi^2}  
I_{\rm f}\Big(\frac {l_0}{T \!+\! T^\dagger},\frac {l_1}{T \!+\! T^\dagger},1\Big)
(T \!+\! T^\dagger)^{-2}
\end{equation}
and the coefficients defined by eq.~(\ref{corrkahlereff}) are found to be
\begin{equation}
C_{n_0,n_1} = \frac {(-1)^{n_0+n_1}}{4 \pi^2 n_0! n_1!} 
I_{\rm f}^{(n_0,n_1,0)}\Big(\frac {\alpha_0}{T \!+\! T^\dagger},\frac {\alpha_1}{T \!+\! T^\dagger},1\Big)
\frac {(T \!+\! T^\dagger)^{-2}}{(3 M_5^3(T\!+\!T^\dagger))^{n_0+n_1}} \;.
\label{cflat}
\end{equation}
The function $I_{\rm f}(a_0,a_1,b)$ appearing in these equations satisfies the scaling law 
$I_{\rm f}(\gamma a_0,\gamma a_1,\gamma b) = \gamma^{-2} I_{\rm f}(a_0,a_1,b)$,
and is defined as
\begin{equation}
I_{\rm f}(a_0,a_1,b) = - \int_0^\infty dx\, x\, {\rm ln} 
\bigg[1 - \frac{1 - a_0\,x/2}{1 + a_0\,x/2}\,
\frac{1 - a_1\,x/2}{1 + a_1\,x/2}\, e^{- b x}\bigg] \;.
\label{fflat}
\end{equation}
The effect of the localized kinetic terms is parametrized by the dimensionless variables $\epsilon_i = \alpha_i/(T \!+\! T^*)$. 
For $\epsilon_i = 0$, all the coefficients are positive and proportional to $c_{\rm f} = I_{\rm f}(0,0,1) \approx 1.202$. 
More precisely, for the first four operators, the functional factor $(-1)^{n_0+n_1}I_{\rm f}^{(n_0,n_1,0)}$ is found to be 
$c_{\rm f}$, $2 c_{\rm f}$, $2 c_{\rm f}$ and $6 c_{\rm f}$. For $\epsilon_i \neq 0$, the situation changes in an interesting 
way \cite{Rattazzi:2003rj}. In particular, for small $\epsilon_0$ and large
$\epsilon_1$, these functional factors tend to $-(3/4)c_{\rm f}$, $-(3/2)c_{\rm f}$, 
$(4 \ln 2/\zeta(3))\, \epsilon_{\rm h}^{-2}\,c_{\rm f}$ and $(4 \ln 2/\zeta(3))\, \epsilon_{\rm h}^{-2}\,c_{\rm f}$.
The reflected situation, with large $\epsilon_0$ and small $\epsilon_1$, is perfectly similar.

In this case, the two branes are completely equivalent. For concreteness, let us put the visible 
sector at $y_0$ and the hidden sector at $y_1$, that is ${\rm v}=0$ and ${\rm h}=1$ in our notation, 
and take $\epsilon_0=0$ and $\epsilon_1 = \epsilon_{\rm h}$. We have then $M_{\rm KK} = 2 (T \!+\! T^*)^{-1}$ 
and $M_{\rm P}^2 = M_5^3 (T\!+\!T^*)(1 + \epsilon_{\rm h})$. The radion-mediated and brane-mediated 
contributions to the soft scalar squared masses are given respectively by $- \partial_T \partial_{T^\dagger} C_{1,0}|F_T|^2$ 
and $- C_{1,1} |F_{\rm h}|^2$. After rewriting the derivatives with respect to $T\!+\! T^\dagger$ as 
derivatives with respect to the last argument of the function (\ref{fflat}), we find:
\begin{eqnarray}
m_{0}^2 &=& b\,\alpha_{\rm gau}^2\, |F_S|^2
+ \Big[\frac {1}3 (1 + \epsilon_{\rm h})\, I_{\rm f}^{(1,0,2)}(0,\epsilon_{\rm h},1)\Big]\,
\alpha_{\rm gra}\,\frac {|F_T|^2}{(T\!+\!T^*)^2} \nn \\
&\;& +\,\Big[\!-\! \frac {1}9 (1 + \epsilon_{\rm h})^2 I_{\rm f}^{(1,1,0)}(0,\epsilon_{\rm h},1) \Big]\,
\alpha_{\rm gra}\, \frac {|F_{\rm h}|^2}{M_{\rm P}^2} \;.
\label{m02}
\end{eqnarray}
For $\epsilon_{\rm h}$ much larger than $1$, the numerical coefficient in the first bracket becomes positive and 
grows like $\epsilon_{\rm h}$, whereas the one in the second stays negative and of order $1$. The possibility of 
achieving a satisfactory result depends however on the dynamics of the model, through the values of 
$F_S$, $F_T/(T\!+\!T^*)$ and $F_{\rm h}/M_{\rm P}$.

A simple class of viable models is obtained by considering the following effective superpotential, which can arise 
for example through gaugino condensation \cite{stabflat}:
\begin{equation}
W = \Lambda_a^3 e^{-n \Lambda_a T} + \Lambda_b^3 + M_{\rm susy}^2\, \Phi_{\rm h} \;.
\end{equation}
In the limit $\Lambda_a \gg \Lambda_b$, there is a stable solution with large 
$T \sim [n\Lambda_a \ln (\Lambda_a/\Lambda_b)]^{-1}$, with a cosmological constant
that can be made to vanish by tuning one of the scales as a function of the others.
The solution is then a function of two parameters, which can be taken to be the supersymmetry
breaking scale $M_{\rm susy}$ and the large dimensionless radius parameter $t = n \Lambda_a T$. 
In terms of these quantities, the $F$-terms are given by
\begin{equation}
F_S \sim \frac {M_{\rm susy}^2}{M_{\rm P}} \;,\;\;
\frac {F_T}{T\!+\!T^*} \sim \frac 1t \frac {M_{\rm susy}^2}{M_{\rm P}} \;,\;\;
\frac {F_{\rm h}}{M_{\rm P}} \sim \frac {M_{\rm susy}^2}{M_{\rm P}} \;.
\end{equation}
In this situation, the sum of the two radion-mediated and brane-mediated gravitational contributions 
becomes positive for $\epsilon_{\rm h} \sim t^2$, and competes with the anomaly-mediated contribution 
if $\epsilon_{\rm h} t^{-2} \alpha_{\rm gra} \sim \alpha_{\rm gau}^2$, that is 
$M_{\rm KK}/M_{\rm P} \sim 4 \pi \alpha_{\rm gau}$ as very generically estimated in eq.~(\ref{est}).
In this situation, we get flavour-universal and non-tachyonic soft masses given by
$m_0 \sim m_{1/2} \sim \alpha_{\rm gau}\,m_{3/2}$. The radion mass is found to be of comparable 
magnitude: $m_{\rm radion} \sim m_{3/2}$.

\section{Strongly warped models}

In the strong warping limit $k R \gg 1$, it is convenient to parametrize the radion superfield by
$\omega = e^{-kT}$. The effective K\"ahler function can then be organized as a series in powers 
of the small parameter $\omega\omega^\dagger$. The tree-level part is given by
\begin{equation}
\Omega = - 3 M_5^3 \big(k^{-1} + l_0 \big) + 3 M_5^3 \big(k^{-1} - l_1 \big) \omega \omega^\dagger \;.
\end{equation}
Similarly, retaining only the leading term, the one-loop correction (\ref{result}) simplifies to
\begin{equation}
\Delta \Omega = \frac {1}{4 \pi^2}  I_{\rm w}(l_0 k,l_1 k)\, (k\, \omega \omega^\dagger)^2
\end{equation}
and the coefficients defined by eq.~(\ref{corrkahlereff}) to
\begin{equation}
C_{n_0,n_1} = 
\frac {(-1)^{n_0+n_1}}{4 \pi^2 n_0! n_1!}I_{\rm w}^{(n_0,n_1)}(\alpha_0 k,\alpha_1 k)
\frac {(k\, \omega \omega^\dagger)^2}{(3 M_5^3 k^{-1})^{n_0+n_1}} \;.
\label{cwarped}
\end{equation}
The function $I_{\rm w}(a_0,a_1)$ appearing in these expressions in defined by
\begin{equation}
I_{\rm w}(a_0,a_1) = \frac {\pi}4 \int_0^\infty dx\, x^3\, 
\frac {\hat K_1(x)}{\hat I_1(x)} \frac 1{1 + a_0} 
\frac {1 - a_1\,x/2\, \hat K_2(x)/\hat K_1(x)}{1 + a_1\, x/2\, \hat I_2(x)/\hat I_1(x)} \;.
\label{fwarp}
\end{equation}
The effect of the localized kinetic terms is parametrized in this case by $\epsilon_i = \alpha_i k$. 
For $\epsilon_i = 0$, all the $C_{n_0,n_1}$'s are positive and proportional to 
$c_{\rm w} = I_{\rm w}(0,0) \approx 1.165$.
More precisely, for the first four operators, the factor $(-1)^{n_0+n_1}I_{\rm w}^{(n_0,n_1,0)}$
is found to be $c_{\rm w}$, $c_{\rm w}$, $2 c_{\rm w}$ and $2 c_{\rm w}$.
For $\epsilon_i \neq 0$, the situation again becomes more interesting \cite{Gregoire:2004nn}.
For small $\epsilon_0$ and sizeable $\epsilon_1$, $C_{0,0}$ and $C_{1,0}$ become negative, 
whereas $C_{1,0}$ and $C_{1,1}$ stay positive. For large $\epsilon_0$ and small $\epsilon_1$, 
on the other hand, all the coefficients stay positive and get just suppressed. 

In this case, the two branes are not equivalent and there are two possible
kinds of models. We will however consider only the case where the visible sector is at $y_0$ 
and the hidden sector at $y_1$, that is ${\rm v}=0$ and ${\rm h}=1$ in our notation, in order to have a 
chance to get positive scalar squared masses, and take $\epsilon_0=0$ and 
$\epsilon_1 = \epsilon_{\rm h}$. We have then $M_{\rm KK} = k|\omega|$ and 
$M_{\rm P}^2 = M_5^3 k^{-1}$. 
The radion-mediated and brane-mediated contributions to the soft scalar squared masses 
are given in this case by $- \partial_{\omega} \partial_{\omega^\dagger} C_{1,0} |F_\omega|^2$ 
and $- C_{1,1} |F_{\rm h}|^2$, and we find:
\begin{eqnarray}
m_{0}^2  &=& b\,\alpha_{\rm gau}^2\, \left|F_S \right|^2
+ \Big[\frac {16}3\, I_{\rm w}^{(1,0)}(0,\epsilon_{\rm h})\Big]\,
\alpha_{\rm gra}\, |F_\omega|^2 \nn \\
&\;& +\,\Big[\!-\! \frac 49 I_{\rm w}^{(1,1)}(0,\epsilon_{\rm h}) \Big]\,
\alpha_{\rm gra}\,\frac {|\omega|^2 |F_{\rm h}|^2}{M_{\rm P}^2} \;.
\label{m02bis} 
\end{eqnarray}
For $\epsilon_{\rm h}$ close to $1$, the numerical coefficient in the first bracket can become positive 
whereas the one in the second remains negative, and both are of order $1$. Again, the possibility of 
achieving a satisfactory result depends however on the dynamics of the model, through the values 
of the parameters $F_S$, $F_\omega$ and $\omega F_{\rm h}/M_{\rm P}$.

A class of viable models can be obtained with the following effective superpotential,
which can be induced for example through gaugino condensation \cite{stabwarp1,stabwarp2}:
\begin{equation}
W = \Lambda_a^3\, \omega^n + \Lambda_{b0}^3 + \Lambda_{b1}^3\, \omega^3
+ M_{\rm susy}^2\, \Phi_{\rm h}\, \omega^3 \;.
\end{equation}
We assume that $\Lambda_{b0}$ is small compared to $\Lambda_{b1}$ but not too small compared to 
$\Lambda_a$, in such a way that we can neglect the $\omega$-dependent term in the negative gravitational 
contribution to the potential. In the limit $\Lambda_a \gg \Lambda_{b1}$ 
with $n > 3$ or $\Lambda_a \ll \Lambda_{b1}$ with $n < 3$, there is a stable solution with small 
$\omega \sim (\Lambda_{b1}/\Lambda_a)^{\frac 3{n-3}}$, with a cosmological constant that can be 
adjusted to zero by tuning again one of the scales.
The solution is then a function of three parameters, which can be chosen to be the supersymmetry
breaking scale $M_{\rm susy}$, the small dimensionless warp factor $\omega$, and one more independent 
dimensionless combination of scales. Omitting the dependence on the latter parameter, that we shall
assume to be generic, the $F$-terms are given by
\begin{equation}
F_S \sim \omega^2 \frac {M_{\rm susy}^2}{M_{\rm P}} \;,\;\;
F_\omega \sim \frac {\omega^2}{1 - \epsilon_{\rm h}}\, \frac {M_{\rm susy}^2}{M_{\rm P}} \;,\;\;
\frac {\omega\,F_{\rm h}}{M_{\rm P}} \sim \omega^2\, \frac {M_{\rm susy}^2}{M_{\rm P}} \;.
\end{equation}
In this situation, the sum of the two radion-mediated and brane-mediated gravitational contributions 
becomes positive for $\epsilon_{\rm h} \sim 1$, and competes with the anomaly-mediated contribution 
if $\alpha_{\rm gra} \sim \alpha_{\rm gau}^2$, which implies again 
$M_{\rm KK}/M_{\rm P} \sim 4 \pi \alpha_{\rm gau}$ as advocated in eq.~(\ref{est}).
In this situation, we get flavour-universal and non-tachyonic soft masses given by
$m_0 \sim m_{1/2} \sim \alpha_{\rm gau}\,m_{3/2}$. The radion mass is in this case found to be of 
even larger magnitude: $m_{\rm radion} \sim \omega^{-1} m_{3/2}$.

\section{Conclusions}

The results of the previous sections show that minimal sequestered models can become viable, with gaugino 
masses dominated by anomaly mediation and scalar squared masses dominated by radion mediation. This 
requires however the presence of a large localized kinetic term for the bulk gravitational fields in the hidden 
sector, with a value that is close to its warping-dependent maximal value. It would be interesting to further 
investigate this extreme situation, and in particular to understand whether it can occur in any natural way and to
what extent subleading flavour-violating effects are suppressed.

\section*{Acknowledgments}

It is a pleasure to thank my collaborators in refs.~\cite{Rattazzi:2003rj} and \cite{Gregoire:2004nn}, on 
which this review is based. This work has been partly supported by the Swiss National Science Foundation 
and by the Commission of the European Communities under contract MRTN-CT-2004-005104.

\end{document}